\documentclass{jaa}
\usepackage{txfonts}
\usepackage{graphicx}
\newcommand{\newblock}{}
\usepackage[authoryear]{natbib}
\begin{document}
\title[Microvariability Detection of Mrk 421]{Microvariability Detection of Mrk 421}
\author[X. Chen, S.M. Hu and D.F. Guo]%
       {Xu Chen, ShaoMing Hu\thanks{corresponding author, e-mail:husm@sdu.edu.cn} and DiFu Guo\\
       School of Space Science and Physics, Shandong University, Weihai\\180 Cultural West Road, Weihai, Shandong 264209, China}
\maketitle
\label{firstpage}
\begin{abstract}
The BL Lac object Mrk 421 was observed in optical bands from 2009 April to 2012 May with the 1.0 m telescope at Weihai Observatory of Shandong University. Microvariability was analysed by C and F tests, but no significant microvariability was detected during our observations.
\end{abstract}
\section{Introduction}
The BL Lac object Mrk 421(B2 1101+384) is one of the closest blazars, with a redshift of z=0.031. Mrk 421 was the first BL Lac object detected at $\gamma$-Ray energy range\citep{lin92} and was classified as a high frequency peaked blazar (HBL). It has been observed extensively in optical bands and it was believed to be strongly variable in optical bands \citep{miller75,liu97}. Microvariability is regarded as an effective way to investigate properties of Active Galactic Nuclei (AGNs). In this work, we tried to detect microvariability of Mrk 421, but no microvariability was detected during our observations.
\section{Observations and Data Reduction}
From 2009 April to 2012 May, 2670 observations of Mrk 421 was obtained using the 1.0 m Cassegrain telescope at Weihai Observatory of Shandong University equipped with a PIXIS 2048B CCD camera. The scale of the CCD is 0.35$''$, and the field of view is about 12$'$ $\times$ 12$'$. The seeing usually ranged from 1.5$''$ to 2.5$''$. The data were processed using an Interactive Data Language (IDL) procedure developed from the NASA IDL Astronomy Library, including image pre-processing and aperture photometry. V and R magnitudes were derived using differential photometry with an aperture of 6.35$''$. Stars 1 and 2 or 2 and 3 \citep{villata98} were used as the standard star and the check star. After data filtering, we derived 2352 valid data points.
\section{Results and Conclusions}
Microvariability of Mrk 421 was investigated by the commonly used quantity C \citep{diego10}, which was defined as $C={{\sigma_{Q}}\over{\sigma_C}}$, where $\sigma_{Q}$ and $\sigma_C$ are the standard deviation of the source and the check star. 
The F test, which is regarded as a proper statistic for photometry \citep{Gaur2012a}, was also used to detect microvariability. F is defined as $F={{S^{2}_{Q}}\over{S^{2}_{C}}}$, where $S^{2}_{Q}$ is the variance of the source  and $S^{2}_{C}$ is that of the check star. 
 Here we take $1\%$ as the significance level 
for both C and F tests.\par
\begin{figure}[!h]
\centering
\includegraphics[keepaspectratio=1,width=11cm, angle=0, trim=0 45 0 50 ,clip=true]{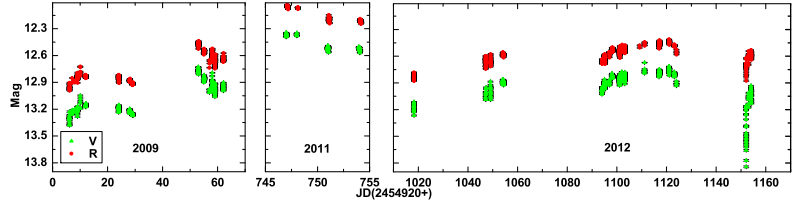}
\caption{Light curves of Mrk 421 between 2009 April and 2012 May}
\label{fig:lc}
\end{figure}

Brightness of Mrk 421 changed 1.50 ($13.84-12.34$) and 0.96 ($12.99-12.03$) magnitudes 
in V and R band, respectively. Light curves are shown in Figure \ref{fig:lc}.
However, no significant microvariability was detected using C and F tests on 37 nights for 73 light curves, whose
data points are more than 5 within one night. Our results are in accordance with results by \cite{Gaur2012a}. Previous works indicated that HBLs show less variability than low frequency peaked blazars \citep{heidt96,heidt98,Gaur2012b}, which was supported by our results. The scenario of stronger magnetic fields in HBLs \citep{romero99} is reasonable for explaining this phenomenon.

\noindent{{\textbf{Acknowledgements:}}} \\
\noindent{This research was partially supported by NSFC under grant Nos. 11143012, 11203016, 10778619 and 10778701, and by the NSF of Shandong Province under grant No. ZR2012AQ008.

\label{lastpage}
\end{document}